\def\kms{\relax \ifmmode {\,\rm km\,s}^{-1}\else \,km\,s$^{-1}$\fi}

\def\mincir{\ \raise-2.truept\hbox{\rlap{\hbox{$\sim$}}\raise5.truept
    \hbox{$<$}\ }}
\def\magcir{\ \raise-2.truept\hbox{\rlap{\hbox{$\sim$}}\raise5.truept
    \hbox{$>$}\ }}
\def\gr{$^\circ$}
\def\sm{M$_\odot$}

\def\arcsec{\hbox{$^{\prime\prime}$}}
\def\nii{[N {\sc ii}]}
\def\sii{[S {\sc ii}]}

\def\oiii{[O {\sc iii}]}

\def\ha{H$\alpha$}
\def\hb{H$\beta$}
\def\chb{$c_{\rm H\beta}$}

\def\ga{\mathrel{\mathchoice {\vcenter{\offinterlineskip\halign{\hfil
$\displaystyle##$\hfil\cr>\cr\sim\cr}}}
{\vcenter{\offinterlineskip\halign{\hfil$\textstyle##$\hfil\cr>\cr\sim\cr}}}
{\vcenter{\offinterlineskip\halign{\hfil$\scriptstyle##$\hfil\cr>\cr\sim\cr}}}
{\vcenter{\offinterlineskip\halign{\hfil$\scriptscriptstyle##$\hfil\cr>\cr\sim\cr}}}}}
\documentclass[longauth]{aa}
\usepackage{txfonts}
\usepackage{graphicx}
\input{epsf}
\begin{document}
\title{The ``Pr\'\i ncipes de Asturias'' nebula: a new quadrupolar planetary
nebula from the IPHAS survey\thanks{The object, whose official name is IPHASX
J012507.9+635652, is called as ``Nebulosa de los Pr{\'{\i}}ncipes de
Asturias'' after its
dedication by the Instituto de Astrof{\'{\i}}sica de Canarias to the Spanish
Princes on the occasion of
their wedding, which took place in Madrid on May, 22nd, 2004}}
\author {A. Mampaso \inst{1}  R.L.M. Corradi \inst{2,1}  K. Viironen \inst{1}
P. Leisy \inst{2,1}
R. Greimel \inst{2} J.E. Drew \inst{3}  M.J. Barlow \inst{4} D.J. Frew \inst{5}
J. Irwin \inst{6} R.A.H. Morris \inst{7} Q.A. Parker \inst{5}
S. Phillipps \inst{7} E.R. Rodr{\'{\i}}guez-Flores \inst{8,1} \and A.A.
Zijlstra \inst{9}
}
   \offprints{A. Mampaso}
   \institute{   Instituto de Astrof{\'{\i}}sica de Canarias, 38200 La Laguna,
Tenerife, Spain.\\
		\email{amr@iac.es; kerttu@iac.es}
         \and
 		Isaac Newton Group. Ap.\ de Correos 321, 38700 Sta. Cruz de la Palma,
Spain.\\
		\email{rcorradi@ing.iac.es; pleisy@ing.iac.es; greimel@ing.iac.es}
 	\and
		Imperial College, Blackett Laboratory, Exhibition Road, London, SW7 2AZ,
UK.\\
		\email{j.drew@imperial.ac.uk}
	\and
		University College London. Department of Physics and Astronomy. Gower St.
London WC1E 6BT, UK.\\
		\email{mjb@star.ucl.ac.uk}
	\and
		Department of Physics, Macquarie University, NSW 2109, Australia.\\
	    	\email{dfrew@ics.mq.edu.au; qap@ics.mq.edu.au}
	\and
		Cambridge Astronomical Survey Unit, Institute of Astronomy, Cambridge, UK.\\
		\email{jmi@ast.cam.ac.uk}
	\and
		Astrophysics Group, Department of Physics, Bristol University, Tyndall
Avenue, Bristol, BS8 1TL, UK.\\
		\email{r.morris@bristol.ac.uk; s.phillipps@bristol.ac.uk}
	\and
		Instituto de Geof{\'{\i}}sica y Astronom{\'{\i}}a. Calle 212, No. 2906, CP
11600, La Habana, Cuba.\\
		\email{erflores@iga.cu}
	\and
		School of Physics and Astronomy, Manchester Univ., Sackville Street, P.O. Box
88, Manchester M60 1QD, UK.\\
	    	\email{aaz@iapetus.phy.umist.ac.uk}
		  }
   \date{Received December, 26th, 2005; accepted July, 10th, 2006 }
\abstract{The Isaac Newton Telescope Photometric H$\alpha$ Survey (IPHAS) is
currently mapping the Northern Galactic plane reaching to $r'$=20 mag with
typically 1\arcsec\ resolution. Hundreds of Planetary Nebulae (PNe), both
point-like and resolved, are expected to be discovered. We report on the
discovery of the first new PN from this survey: it is an unusual object
located at a large galactocentric distance and has a very low oxygen
abundance.}  {Detecting and studying new PNe will lead to improved estimates
of the population size, binary fraction and lifetimes, and yield new insights
into the chemistry of the interstellar medium at large galactocentric
distances.}  {Compact nebulae are searched for in the IPHAS photometric
catalogue, selecting those candidates with a strong H$\alpha$ excess in the
$r'-H\alpha$ vs. $r'-i'$ colour-colour diagram. Searches for extended nebulae
are by visual inspection of the mosaics of continuum-subtracted H$\alpha$
images at a spatial sampling of 5$\times$5~arcsec$^2$. Follow-up spectroscopy
enables confirmation of the PNe, and their physico-chemical study} {The first
planetary nebula discovered via IPHAS imagery shows an intricate morphology:
there is an inner ring surrounding the central star, bright inner lobes with
an enhanced waist, and very faint lobular extensions reaching up to more than
100$''$. We classify it as a quadrupolar PN, a rather unusual class of
planetary showing two pairs of misaligned lobes.  From long-slit spectroscopy
we derive $T_e$[N{\sc ii}] =12800$\pm$1000 K, $N_e$ = 390$\pm$40 cm$^{-3}$,
and chemical abundances typical of Peimbert's Type I nebulae ($He/H$ =0.13,
$N/O$ =1.8) with an oxygen abundance of $12+log(O/H)$=8.17$\pm$0.15.  A
kinematic distance of 7.0$^{+4.5}_{-3.0}$~kpc is derived, implying an
unusually large size of $>$4 pc for the nebula.  The photometry of the central
star indicates the presence of a relatively cool companion.  This, and the
evidence for a dense circumstellar disk and quadrupolar morphology, all of
which are rare among PNe, support the hypothesis that this morphology is
related to binary interaction.  } {}
\keywords{planetary nebula --
                morphology --
                chemical abundances}
\titlerunning {A new quadrupolar planetary nebula}
\authorrunning {A. Mampaso et al.}
\maketitle

\section{Introduction}
The INT/WFC Photometric H$\alpha$ Survey of the Northern Galactic Plane (IPHAS)
is an ambitious programme supported by an international collaboration among 15
institutes. The survey aims at producing a
complete and detailed H$\alpha$ map of the Galactic Plane, within the latitude
range, $-5^\circ \le\ b^{II}\le\ +5^\circ $, north of the celestial equator.
The survey will cover a total of 1800 square degrees of sky. It started in
August 2003, with a target completion date of the end of 2006. It is estimated
to take a total of 30 observing weeks (mostly during bright time),
contributed by all three national communities involved --
the UK, Spain and the Netherlands.

IPHAS makes use of the Wide Field Camera (WFC) of the 2.5m Isaac
Newton Telescope (INT) at the Observatorio del Roque de los Muchachos
on La Palma, Spain. A narrow-band H$\alpha$ filter ($\lambda_c$ = 6568 {\AA};
$FWHM$ = 95 {\AA}) and two broad-band Sloan $r'$, $i'$ filters are used for
matched 120, 30, and 10 s exposures, respectively, spanning the
magnitude range $13\leq r' \leq 20$ for point sources. The survey area is
covered in double pass, such that every pointing is repeated at an offset
of 5~arcmin in both right ascension and declination.  Pipeline data
reduction and data distribution are handled by the Cambridge
Astronomical Survey Unit (CASU {\it http://archive.ast.cam.ac.uk/}).

IPHAS is the first fully-photometric H$\alpha$ survey of the Galactic
plane, and it will complement the recently completed photographic survey
of the southern Galactic plane performed with the AAO UK Schmidt
Telescope (\cite{parkera}) {\it
http://www-wfau.roe.ac.uk/sss/halpha/index.html}.

IPHAS is expected to discover up to 50000 new emission-line stars,
including young stars (T Tau, Herbig AeBe stars, etc.), and evolved ones
(post-AGB, LBVs, etc.), as well as different classes of interacting
binaries (CVs, symbiotic stars, etc.) in addition to thousands of
extended nebulae such as planetary nebulae, HII regions, SN
remnants, H-H objects, etc.  Further information on the
objectives of IPHAS, its products, and early scientific results can be
found in \cite{drew} and in the public web site at
{\it http://www.iphas.org/}.

\subsection{IPHAS and planetary nebulae}
One significant contribution of IPHAS will be in the field of planetary
nebulae (PNe). The AAO-UKST Southern photographic \ha\ survey picked up nearly
1000 new PNe (\cite{parkerb}). IPHAS has a similar detection limit for
extended objects but an improved spatial resolution, and it is estimated that
IPHAS will discover several hundred new PNe, including very low
surface-brightness objects and faint compact nebulae. The lower nebular
surface brightnesses, relative to already known PNe, will make studies of
their central stars easier to carry out, ultimately enabling much better
statistics to be obtained on the overall incidence of binarity (currently a
major unknown; c.f. \cite{orsola}) and the frequency of different binary
types. The two surveys together will also allow us to obtain better estimates
of the PN population size in the solar vicinity and throughout the disc of the
Galaxy, a figure which is intimately related to our understanding of the PN
density per unit luminosity in spiral galaxies, the stellar death rate, PNe
lifetimes, and their absolute luminosities (\cite{buzzoni}).

A number of the PNe discovered by IPHAS will be located along very
interesting lines-of-sight, such as the direction of the Galactic
anticentre, providing new targets to probe stellar properties and the
chemistry of the interstellar medium at large galactocentric distances.
For these reasons, a significant effort is being made, within the IPHAS
collaboration, to perform a systematic search for ionized
nebulae, including PNe. Two search techniques are used.

First, candidate compact nebulae are selected from the photometric catalogue
that is automatically created by the IPHAS data reduction pipeline.  Both
stellar and quasi-stellar sources are detected, and we select those
with a strong H$\alpha$ excess in the IPHAS $r'-H\alpha$ vs. $r'-i'$
colour-colour diagram (see Drew et al. 2005, and \cite{corradi}).
This allowed us to select 66 candidates in the data analyzed so far,
corresponding to the observations obtained from 2003 to July 2004.

Second, extended nebulae which are not detected by the automated
photometry are found by visual inspection of the IPHAS images.  In
particular, the CASU web interface includes analysis software which
produces mosaics of continuum-subtracted H$\alpha$ images at any
spatial sampling. We adopt a spatial binning of 15$\times$15~pixels of
the WFC CCDs, corresponding to 5$\times$5~arcsec$^2$, which allows us
to get $S/N \sim$4 at $F(H\alpha )$ = 2 10$^{-17}$ erg  arcsec$^{-2}$
cm$^{-2}$ s$^{-1}$ in each rebinned spatial element. Preliminary analysis
of a few square degrees led to the detection of 20 candidate PNe. An article
with further details on the search methods will be published as
soon as the scanning of a significant fraction of the IPHAS area is
completed in a systematic way for both small and large nebulae. The
main aim of the present article is to illustrate the kind of results
that we aim to obtain from IPHAS in the field of ionized nebulae, and
its follow-up observations, by presenting the case of the first planetary
nebula discovered by the project.  As well as being first to be found,
this object is also a puzzle.

Spectroscopic and imaging observations for this object are presented in Sect.
2. Data analysis and results are in Sect. 3, which includes a
physico-chemical analysis, a spatio-kinematical model for the inner
nebula, and a discussion of the central star and the distance to the
PN. A general discussion and main conclusions are presented in Sects. 4 and 5,
respectively.

\section{Observations}
\subsection{IPHAS}
\begin{figure*}[ht]
\centering
\includegraphics[width=\textwidth]{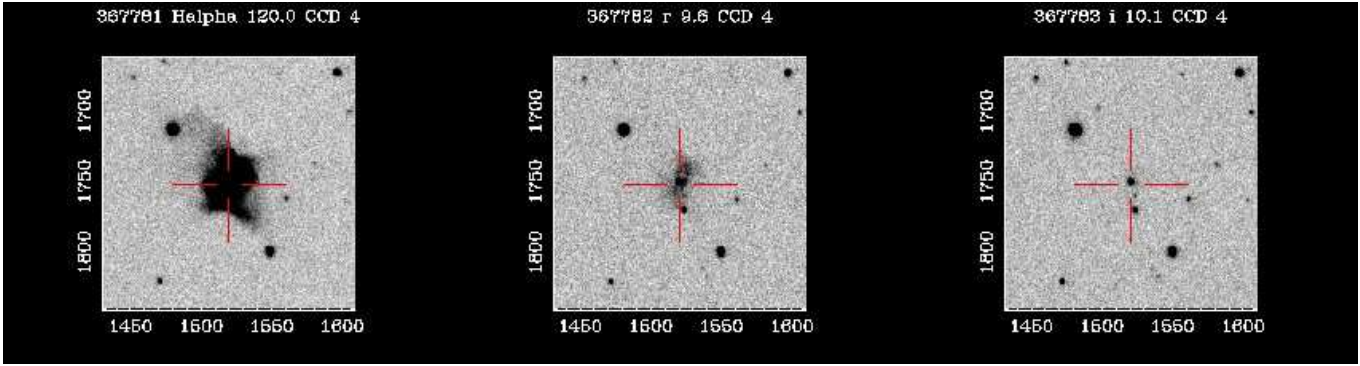}
\caption{Discovery images of the planetary nebula IPHAS PN-1
in \ha, $r'$, and $i'$ with exposures of 120, 10, and 10 seconds, respectively.
Note that the \ha\ filter of IPHAS includes the \nii\ lines at 6548 and 6583
{\AA}
which are more intense than \ha\ in this object (see text).
Images are 1 arcmin on each side; North is up and East to the left.}
\label{iphas}
\end{figure*}

Fig.~\ref{iphas} presents the \ha, $r'$ and $i'$ images of a
nebulous object discovered during the analysis of the IPHAS
photometric catalogue. The images were obtained on October, 13th, 2003 under
seeing conditions of 1.0 arcsec $FWHM$.
The object is located at $RA(2000)$= 1$^h$ 25$^m$ 7.9$^s$; $Dec(2000)$=
+63$^o$ 56$\arcmin$ 52$\arcsec$.  As no sources are catalogued in the
SIMBAD database within 4 arcmin of these coordinates, this is a
genuine new detection of a Galactic nebula. According to the
nomenclature adopted by the collaboration (Drew et al. 2005),
the source was given the name IPHASX J012507.9+635652, where X stands for
`extended'. We will abbreviate to IPHAS PN-1 hereafter.
The nebula is a radio source detected by the NRAO VLA Sky Survey
(\cite{condon}) with an integrated flux of 4.2$\pm$0.5 mJy at 1.4 GHz. It was
not resolved by the radio beam, implying a size smaller than
74$\arcsec\times$44$\arcsec$. On the other hand, no CO nor mid- or far- (MSX
and IRAS) infrared sources are associated with the nebula.

\begin{figure*}
\centering
\includegraphics[width=\textwidth]{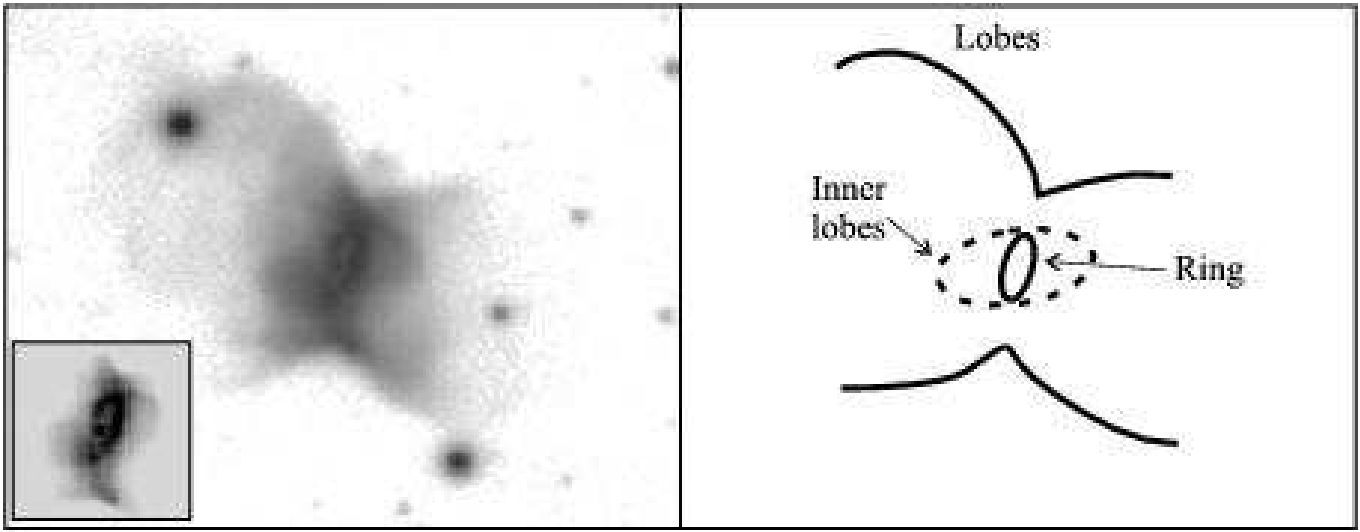}
\caption{({\it Left}) Inner regions of IPHAS PN-1 in \ha\ +\nii\  displayed with a
logarithmic scale. The central star and elliptical ring
are shown in the insert with a linear scale. ({\it Right}) Sketch showing the
central elliptical ring, the main lobes and waist,
and the tilted ellipsoidal inner lobes.}
\label{sketch}
\end{figure*}

The \ha+\nii\ image (Fig.~\ref{iphas} left panel, and Fig.~\ref{sketch})
reveals a complex nebular morphology, consisting of:

-- a relatively bright central star;

-- an inner elliptical ring with semiaxes (at peak) of
3$''$.2$\times$1$''$.3, with the long axis oriented at
$P.A.\sim$160$^\circ$;

-- a pair of incomplete bright lobes. The `waist' of the lobes is at
$P.A.=$170$^\circ$ and has a size of 13$''$, i.e. it is larger than
the ring and tilted with respect to it.  The lobes extend in a
direction which is roughly perpendicular to the waist, show some
deviations from axisymmetry, and decrease abruptly in brightness at a distance
of some 20$''$ from the centre;

-- inside these lobes, apparently arising from the bright inner ring
(but, again, not aligned with it), there is an elliposidal structure
roughly centred on the central star
(possibly another pair of inner lobes/bubbles), whose long axis is
oriented at $P.A.=$100$^\circ$ and has semiaxes of roughly
3$''$.3$\times$6$''$.5.

Fig.~\ref{sketch} shows the inner, brighter region of the nebula
with the different structures labelled.  The presence of nested
bipolar structures is characteristic of a handful of PNe
forming the morphological class of the so-called quadrupolar nebulae
(\cite{manchado}). As this is a rare and puzzling class of PNe,
we decided to take deeper images as well as spectroscopic data of the
newly discovered nebula.

\subsection{Other imaging}
The object was imaged with the INT+WFC on August 4th, 2004 through the
same \ha\ filter (total exposure time 18 min), an \sii\ filter
($\lambda _c$ = 6725 {\AA}; $FWHM$ = 80 {\AA}; exp. time 30~min) and
\oiii\ ($\lambda _c$ = 5008 {\AA}; $FWHM$ = 100 {\AA}; exp. time 18~min)
filter.  Seeing was 1$''$.4 in \ha\ and \oiii\, and 1$''$.0 in
\sii. The nebula is very faint both in \oiii, where emission is
limited to the zone of the ring and inner lobes, and in \sii, where
basically only the ring is visible. As the \ha\ image does not reveal
more details than the original one, this second set of images is not
shown in this paper.

\begin{figure*}
\centering
\includegraphics[angle=-90,width=\textwidth]{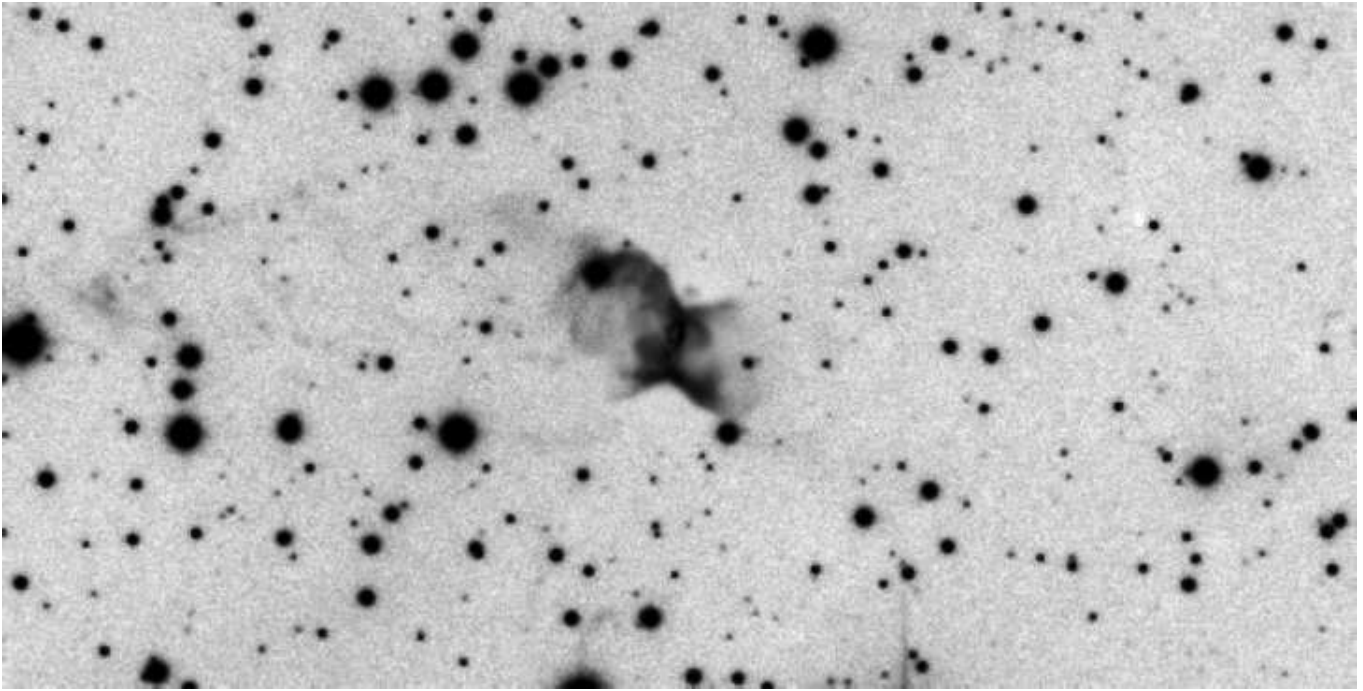}
\caption{IPHAS PN-1 in \ha\ +\nii\  displayed with a logarithmic scale.
The cuts are chosen as to highlight the extended low-level emission. The image
size is 2\arcmin$\times$4\arcmin.}
\label{ha_deep}
\end{figure*}

An additional \ha+\nii\ image with a total exposure time of 1.5~hour
(9$\times$600~s) was obtained with the INT+WFC on September 3rd, 2004. The
night was photometric with seeing of 0.9~arcsec and the image was flux
calibrated using the standard stars GD240 and Feige 110 (\cite{oke}). This
deep image, displayed in Fig.~\ref{ha_deep}, reveals spectacular extensions of
the outer lobes along their Eastern side, with only hints of a corresponding
extension on the opposite side. The tip of the Eastern extension has the shape
of an arrow and is located at a distance of 105~arcsec from the central star,
at $P.A.=83^\circ$, i.e. in an axis perpendicular, within errors, to the
waist. The projected symmetry axes of the different structures in IPHAS PN-1
rotate clockwise from the interior, the ring at $P.A.=$160$^\circ$, to the
exterior, the inner lobes at $P.A.=$100$^\circ$, and the main lobes (including
their faint extensions), at $P.A.=$83$^\circ$.

\subsection{Follow-up spectroscopy}
\subsubsection{High resolution spectroscopy}
IPHAS PN-1 was observed on October, 25th, 2004 with the
2.1m telescope (Observatorio Astron\'omico Nacional, San Pedro
M\'artir, B.C. M\'exico) equipped with the MESCAL echelle spectrograph
and a 90 {\AA} wide \ha\ filter.  The spectral reciprocal dispersion
was 0.1 {\AA}~pix$^{-1}$ over a spectral range from 6543 to 6593 {\AA}.
The slit width and length were 2\arcsec.0 and 5\arcmin.1, respectively.
Exposures of 1200 and 1800 s were taken at $P.A.$=0 and 57$^{\circ}$,
respectively, with median seeing of 1.6\arcsec\ during the night.

\begin{figure}[htbp]
{\centerline {
    \epsfxsize=4.3cm
    \epsffile{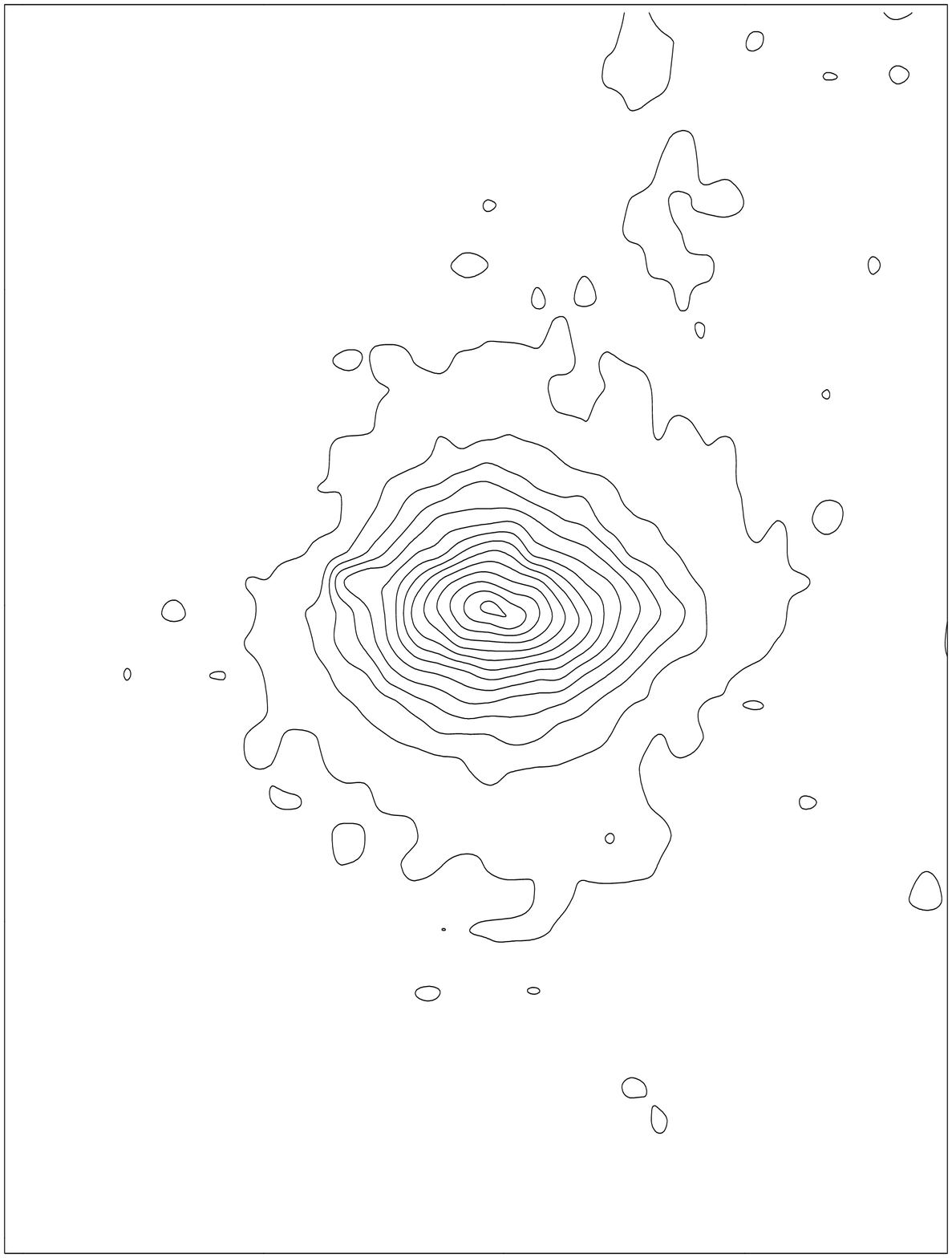}
    \epsfxsize=4.3cm
    \epsffile{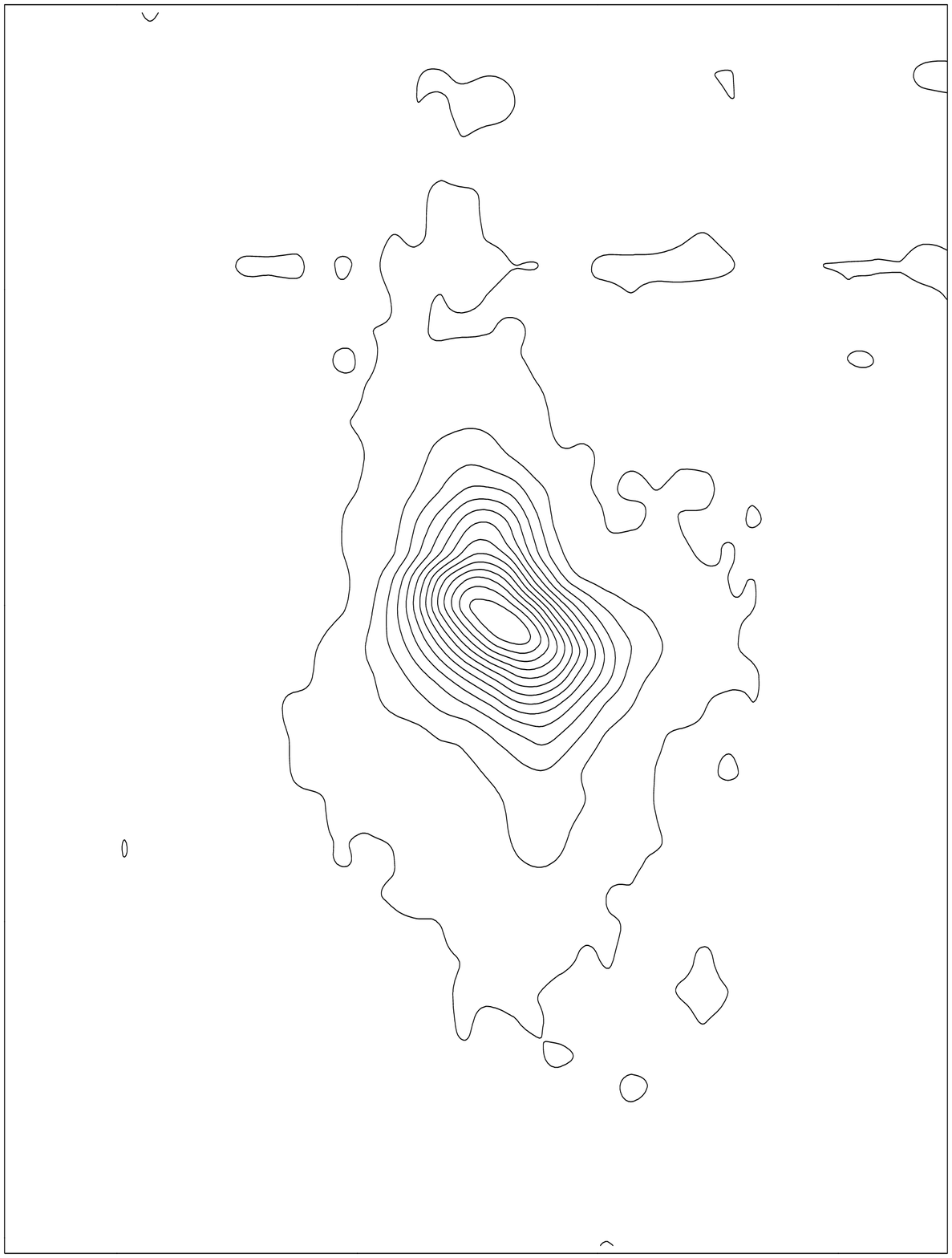}
    }}
{\centerline {
    \epsfxsize=4.3cm
    \epsffile{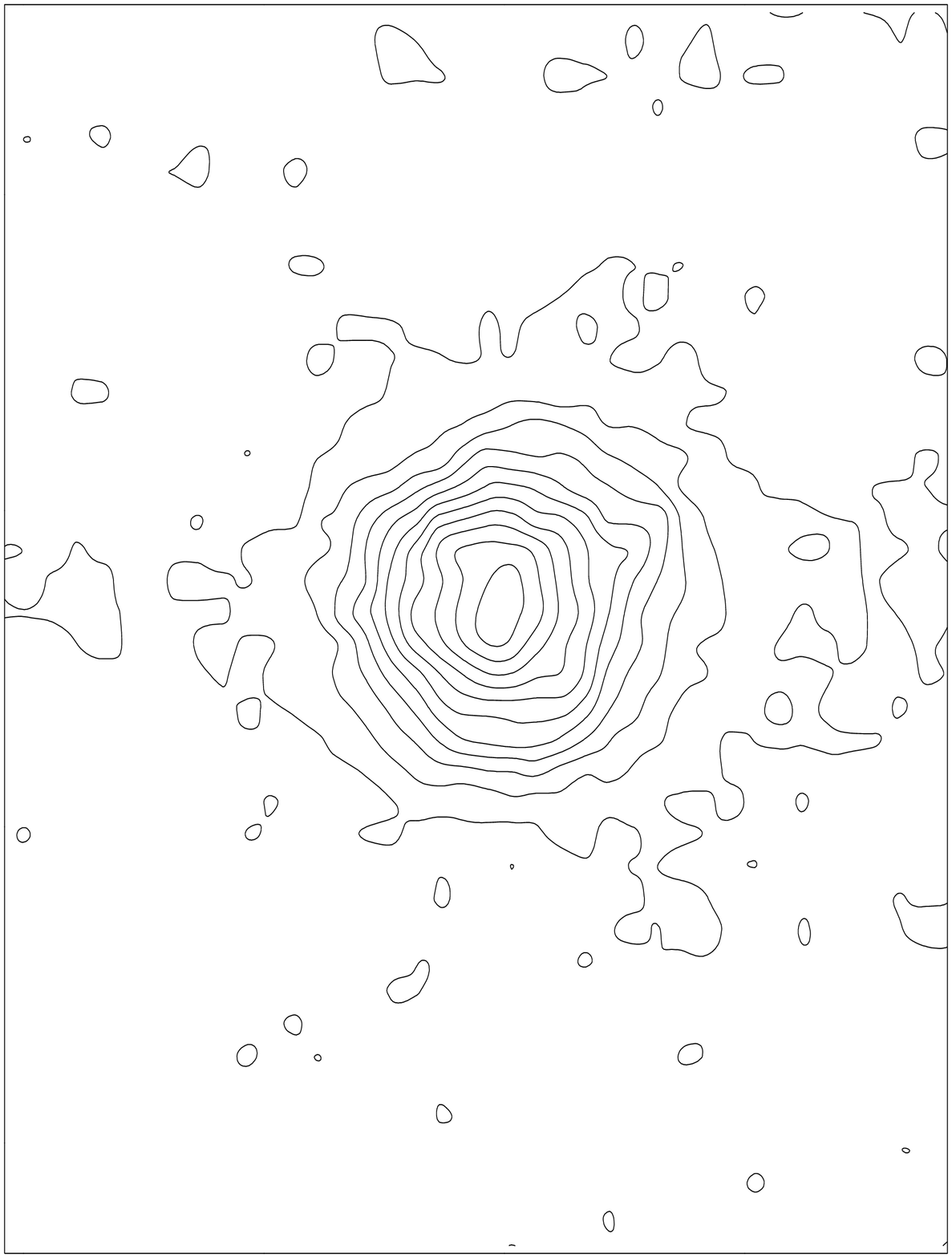}
    \epsfxsize=4.3cm
    \epsffile{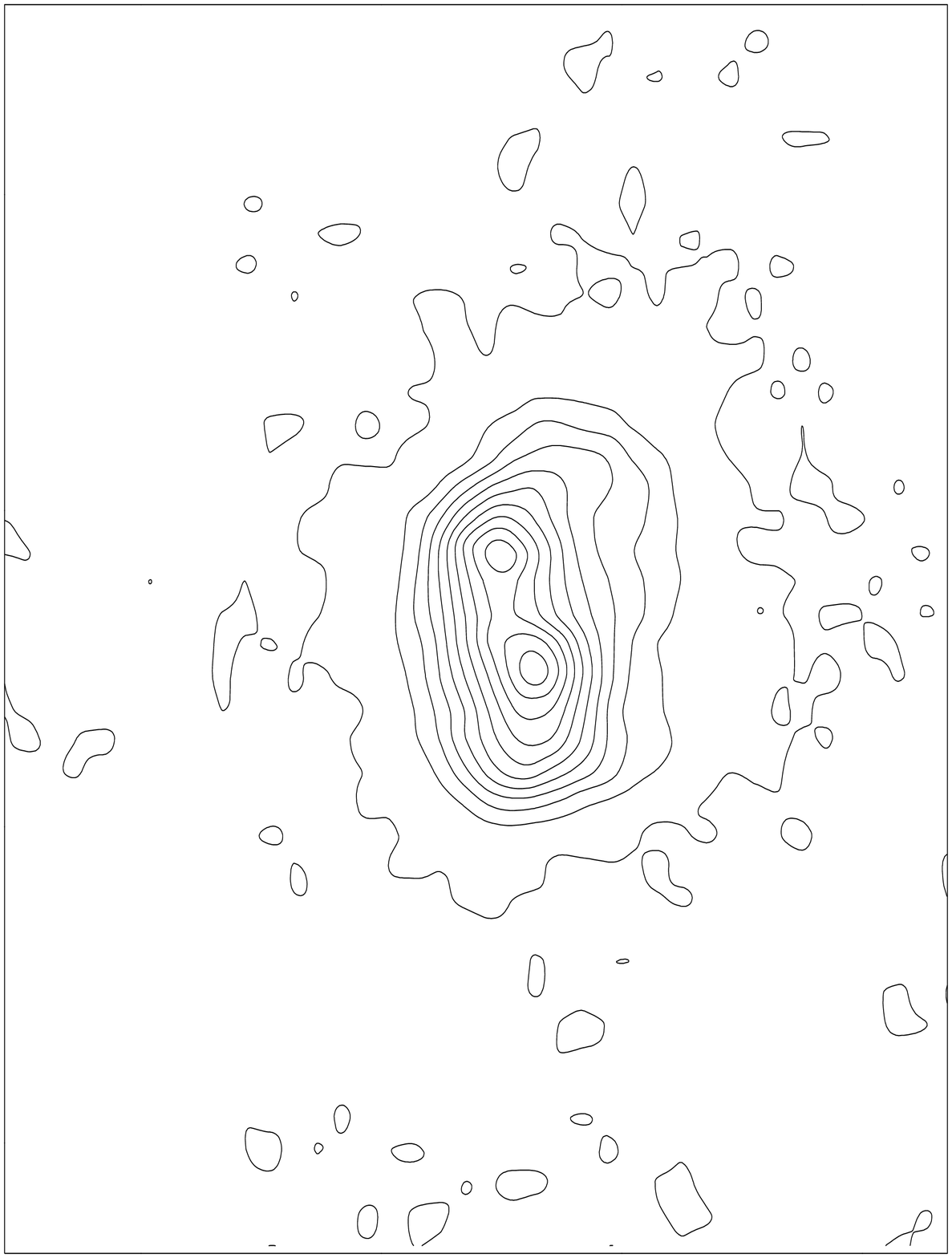}
    }}
 \caption{Position-velocity maps for H$\alpha$ ({\it left panels}) and [NII]
6583 {\AA} ({\it right panels}) from the spectra at $P.A.$= 57$^{\circ}$ ({\it
top panels}) and  $P.A.$= 0$^{\circ}$ ({\it bottom panels}). Contours have
been drawn from the 2-$\sigma$ level to the maximum, linearly spaced and
smoothed by block averaging by 2$\times$2 pixels. The length of the vertical
side of each box corresponds to 48{\arcsec} (North is towards the bottom)
whereas the horizontal side corresponds to 180 km~s$^{-1}$ (red is to the
right).}
\label{spe}
\end{figure}

Four lines were detected: [NII] $\lambda\lambda$ 6548, 6583 {\AA},
He II 6560 {\AA} and \ha . The bright \ha\ and [NII] 6583 {\AA} lines could be
measured along the inner $\sim$20\arcsec\ of the nebula and are shown in
Fig.~\ref{spe}. The rich spatio-kinematic structure seen in the 6583 {\AA}
line, that is presumably present in the \ha\ line, is smeared out by the
higher thermal broadening associated with hydrogen.  These data are analysed
in Section 3.3, and further exploited in Section 3.4.

\subsubsection{Low resolution spectroscopy}
The nebula was observed on September, 23rd and 27th, 2004, with the
4.2m WHT telescope (Observatorio del Roque de los Muchachos, La Palma,
Spain) and the ISIS spectrograph.  Two 1800~s exposures through the
blue arm (Sep. 23rd) plus four 1800~s exposures through both the blue
and red arms (Sep. 27th) were taken using gratings R300B (spectral
dispersion of 0.86 {\AA}~pix$^{-1}$ from 3500 to 6100 {\AA}) and R158R
(dispersion of 1.63 {\AA}~pix$^{-1}$ from 6000 to 10500 {\AA}).  The spatial
scale of the instrument is 0\arcsec.2~pix$^{-1}$ and the slit width
and length were 0\arcsec.9 and 3\arcmin.7, respectively. The slit was
positioned passing through the central star at $P.A.$= 180$^{\circ}$
(Sep. 23rd) and 125$^{\circ}$ (Sep. 27th), matching the parallactic
angle at the middle of each observing period. Seeing was 0\arcsec.8
(Sep. 23th) and 1\arcsec.5 (Sep. 27th). Bias frames, twilight and
tungsten flat-field exposures, arcs and exposures of the standard star
G191-B2B (\cite{oke}) were obtained.  Spectra were reduced and flux
calibrated using the standard IRAF packages for long-slit spectra.
\section{Spectroscopic analysis}
The [OIII] 4363 {\AA} temperature sensitive line was measured with a
$S/N$ ratio of $\sim$30 in the WHT blue spectrum taken at $P.A.$=
180$^{\circ}$ on 23rd September 2004.  It is clearly present in the
spectrum at $P.A.$= 125$^{\circ}$ obtained 4 nights later, but because this
was taken under worse seeing and bright (full moon) sky conditions, its
profile is too noisy to justify a second measurement.
In the following, we will assume for IPHAS PN-1 the $T_e$[O{\sc iii}]
determined from the spectrum of September 23rd, only, adopting a large
uncertainty of 20\% to account for both the formal errors (see below) and
the fact that we implicitly assume an invariant $T_e$[O{\sc iii}] across the
nebula. Otherwise, the physico-chemical analysis described below is based
on the lines measured in the complete (blue+red) spectra of September 27th,
at $P.A.$= 125$^{\circ}$.

\begin{figure*}
\sidecaption
  \includegraphics[width=12cm]{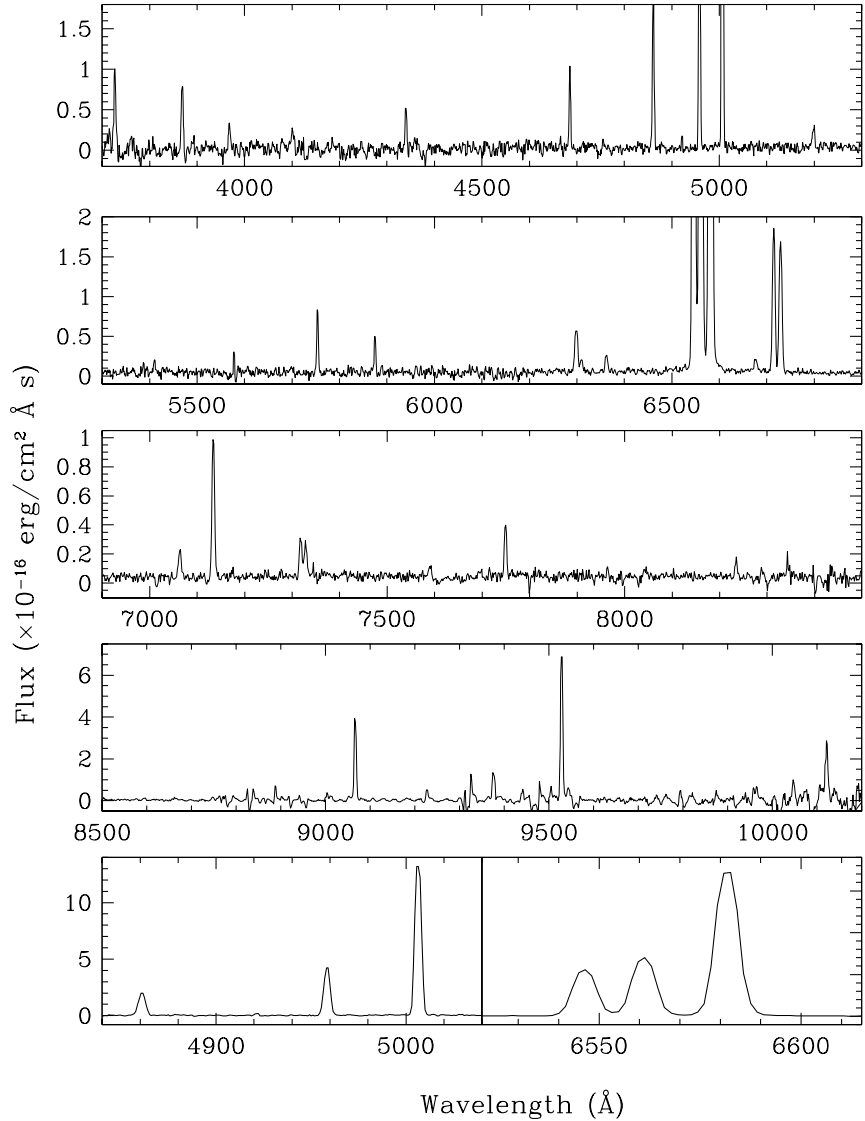}
    \caption{WHT+ISIS spectrum of IPHAS PN-1 at $P.A.$= 125$^{\circ}$ (regions NW
and SE co-added). The lowest panel shows full line profiles for the six
strongest lines, H$\beta$, [OIII] $\lambda$$\lambda$ 4959, 5007 {\AA},
H$\alpha$, and [NII] $\lambda$$\lambda$ 6548, 6583 {\AA}.}
    \label{spec}
\end{figure*}

Three nebular regions were selected: a central one extending
$\pm 0.8$\arcsec\ around the central star, plus two regions of 5.2\arcsec\
located symmetrically with centres at 3.9\arcsec\ NW and SE from the
star. The latter correspond with the two bright sections of the inner ring,
where line emission is stronger. No significant variations of extinction,
density or temperature $T_e$[N{\sc ii}] were found between the NW and SE
regions, and we averaged their spectra to further increase the $S/N$
ratio. The resulting spectrum is shown in Fig.~\ref{spec} and
emission line fluxes are listed in Table ~1. Quoted errors on observed
fluxes include both the statistical poissonian noise and the
systematic contributions of the wavelength and flux calibrations plus
the continuum determination. Fluxes were extinction-corrected by using
\chb\ =2.0$\pm$0.1 (the logarithmic ratio between observed and
dereddened \hb\ fluxes), determined from the observed \ha/\hb\  ratio,
and the reddening law of \cite{cardelli} for $R_V$=3.1.
An average extinction can also be estimated comparing the radio and \ha\
fluxes (c.f. \cite{pottasch}).
The raw integrated \ha +\nii\  flux from the nebula, measured from the image in
Fig.~\ref{ha_deep}, is 1.20$\times10^{-12}$ erg\,cm$^{-2}$s$^{-1}$ and the
correction for \nii\  $\lambda\lambda$6548, 6583 {\AA} emission (using a flux
ratio of [NII]/\ha =3.2 as derived from the spectra; c.f. Table ~1) yields
$F$(\ha )=3.75$\times10^{-13}$ erg\,cm$^{-2}$s$^{-1}$. Comparing this with the
measured radio flux (Sect. 2.1), and assuming that all the extinction is
external to the object and that the \ha/\nii\  ratio is constant along the
nebula, we obtain
\chb\ =1.7$\pm$0.2. This value is in fair agreement with the optical value,
\chb\ =2.0, deduced above, and we will adopt the latter in the following.

\begin{table}
\caption{Observed and dereddened line fluxes for the WHT+ISIS spectrum at
$P.A.$= 125$^{\circ}$ (normalized to \hb\ =100).  Percentage r.m.s. errors are given within brackets.}
 \centering
 \begin{tabular}{@{}lrrr@{}}
 \hline \hline
Line Ident. (\AA)&$\lambda$(Obs.)&Observed flux&Dereddened flux\\
 \hline
{}[O{\sc ii}] 3726.0, 3728.8		     &3726.8    &62.7\ (6.1)   &274.1\ (9.0)\\
{}[Ne{\sc iii}] 3868.7			     &3868.6    &46.0\ (6.0)   &174.8\ (8.5)\\
{}[Ne{\sc iii}] 3967.5 + H$\epsilon$         &3968.0    &18.2\ (12.4)   &61.7\ (13.5)\\
H$\gamma$ 4340.5 			     &4339.9    &25.3\ (6.8)    &51.8\ (7.5)\\
He{\sc ii} 4685.7 		             &4685.3    &47.7\ (4.0)    &59.9\ (4.1)\\
H$\beta$ 4861.3 		             &4860.9    &100.0\ (2.5)   &100.0\ (2.5)\\
{}[O{\sc iii}] 4958.9 		             &4958.4    &194.8\ (1.8)   &173.2\ (1.9)\\
{}[O{\sc iii}] 5006.8 		             &5006.4    &630.3\ (1.6)   &530.3\ (1.7)\\
{}[N{\sc i}] 5197.9, 5200.2 	             &5198.6    &17.3\ (8.5)    &11.8\ (8.7)\\
He{\sc ii} 5411.6 		             &5410.1	&7.9\ (18.4)	&4.4\ (18.6)\\
{}[N{\sc ii}] 5754.6 		             &5753.6    &37.5\ (4.1)    &16.1\ (5.6)\\
He{\sc i} 5875.7 		             &5874.7    &24.7\ (6.5)    & 9.8\ (7.7)\\
{}[O{\sc i}] 6300.3 		             &6298.9    &43.5\ (5.1)    &13.1\ (7.4)\\
{}[S{\sc iii}] 6312.1 		             &6309.6    &11.1\ (21.2)	&3.3\ (21.9)\\
{}[O{\sc i}] 6363.8  		             &6362.3    &14.1\ (14.3)	&4.1\ (15.4)\\
{}[N{\sc ii}] 6548.0 	 	             &6546.4    &876.9\ (1.5)   &226.4\ (6.3)\\
H$\alpha$ 6562.8 		             &6561.0    &1116.0\ (1.4)   &285.6\ (6.3)\\
{}[N{\sc ii}] 6583.4  		             &6581.7    &2732.1\ (1.4)   &690.6\ (6.4)\\
He{\sc i} 6678.1  		             &6676.7    &13.8\ (12.3)	&3.3\ (13.9)\\
{}[S{\sc ii}] 6716.5  		             &6714.9    &141.2\ (2.1)    &32.9\ (6.9)\\
{}[S{\sc ii}] 6730.8  		             &6729.2    &127.4\ (2.2)    &29.5\ (7.0)\\
He{\sc i} 7065.3			     &7063.5    &17.3\ (8.5)     &3.3\ (11.4)\\
{}[Ar{\sc iii}] 7135.8			     &7134.0    &74.7\ (2.8)    &13.5\ (8.2)\\
{}[O{\sc ii}] 7319.4			     &7317.7    &19.8\ (10.3)	&3.2\ (13.2)\\
{}[O{\sc ii}] 7330.3			     &7328.3    &18.6\ (9.7)     &3.0\ (12.8)\\
He{\sc ii} 7592.7			     &7590.0	&6.0\ (26.2)	&0.8\ (27.7)\\
{}[Ar{\sc iii}] 7751.4			     &7749.2    &24.8\ (6.0)     &3.1\ (11.2)\\
He{\sc ii} 8236.8			     &8234.8	&8.2\ (16.4)	&0.8\ (19.6)\\
{}[S{\sc iii}] 9068.9			     &9066.4    &288.3\ (2.3)    &19.0\ (12.5)\\
P9 9229.0				     &9227.5    &37.2\ (10.9)	&2.3\ (16.6)\\
{}[S{\sc iii}] 9531.0			     &9528.5    &524.0\ (2.9)    &29.9\ (13.3)\\
P7  10049.4				     &10047.2	&70.7\ (28.5)	 &3.5\ (31.6)\\
He{\sc ii} 10123.6			     &10121.6   &221.1\ (10.2)	&10.8\ (17.1)\\
\hline
\end{tabular}
\end{table}

\subsection{Physical and chemical properties}
The diagnostic ratios log(\ha/\nii)=$-$0.50 and log(\ha/\sii)=0.66
confirm the PN classification \cite{sabbadin}. Its name, according to
the standard IAU nomenclature, would be \object{PN G~126.6+1.3}, after its
Galactic coordinates.

From the \oiii, \nii, and \sii\ line ratios, electron temperatures of
$T_e$[O{\sc iii}] = 14100$\pm$2800 K, $T_e$[N{\sc ii}] =
12800$\pm$1000 K and a density of $N_e$[S{\sc ii}] = 390$\pm$40
cm$^{-3}$ are obtained.  Ionic and total abundances were calculated
using a two-zone analysis and the NEBULAR IRAF package. Errors were
consistently propagated through the calculations in the way described
by e.g. \cite{perinotto98}. Chemical abundances are shown in Table~2;
associated errors include the uncertainties in the extinction,
temperature, and the measured ionization correction factors ({\it
icf}), in addition to the errors in the line fluxes.  However, no
uncertainties in the assumed extinction law nor in the {\it icf}
scheme are considered. The latter was adopted from Kingsburgh \&
Barlow (1994; see also \cite{perinotto04},
in particular for a discussion of the Sulphur abundance determination).

\begin{table*}
\caption{Ionic and total abundances of IPHAS PN-1.
Percentage {\it {r.m.s.}} errors are given within brackets. ($^a$) Abundances
for Type I PNe from \cite{kb94}. ($^b$) Abundances for the three Type
I PNe with $D_{GC}\ge 11$~kpc from \cite{costa}. 1-$\sigma$ dispersions of both samples are given within square brackets}
\centering
 \begin{tabular}{lrrrr}
 \hline \hline
Ion/Element&Abundance&12+log(X/H)&Type I$^a$&Type I$^b$\\
 \hline
 He$^+$/H &  0.076\ (8)  & & & \\
 He$^{2+}$/H  &  0.054\ (7)  & & & \\
 {\bf He/H } & {\bf 0.130\ (11)} & {\bf 11.11$\pm$0.05} & 11.11[$\pm$0.016] &
11.22[$\pm$0.015] \\
 \hline
 O$^0$/H &   1.06E-05\ (23) & & & \\
 O$^+$/H  &   4.00E-05\ (27) & & & \\
 O$^{2+}$/H  &   6.48E-05\ (49) & & & \\
 {\it icf}(O)  &   1.42\ (13) & & & \\
 {\bf O/H}  &   {\bf 1.49E-04\ (34)} & {\bf 8.17$\pm$0.15} & 8.65[$\pm$0.15] &
8.46[$\pm$0.17] \\
 \hline
 N$^+$/H  &   7.39E-05\ (17) & & & \\
 {\it icf}(N)  &   3.73\ (44) & & & \\
 {\bf N/H} &   {\bf 2.76E-04\ (46)} & {\bf 8.44$\pm$0.20} & 8.72[$\pm$0.15] &
8.50[$\pm$0.12] \\
 \hline
 Ne$^{2+}$/H &   5.68E-05\ (59) & & & \\
 {\it icf}(Ne) &   2.30\ (60) & & & \\
 {\bf Ne/H} &   {\bf 1.31E-04\ (84)} &  {\bf 8.12$\pm$0.36} & 8.09[$\pm$0.15]
&  \\
 \hline
 S$^+$/H  &   8.81E-07\ (18) & & & \\
 S$^{2+}$/H  &   2.57E-06\ (65) & & & \\
 {\it icf}(S) &   1.18\ (39) & & & \\
 {\bf S/H} &   {\bf 4.07E-06\ (62)} &  {\bf 6.61$\pm$0.27} & 6.91[$\pm$0.30]  &
6.55[$\pm$0.25] \\
 \hline
 Ar$^{2+}$/H  &   5.83E-07\ (35) & & & \\
 {\it icf}(Ar) &   1.87\ (21) & & & \\
 {\bf Ar/H} &   {\bf 1.09E-06\ (41)} & {\bf 6.04$\pm$0.18} & 6.42[$\pm$0.30]  &
6.54[$\pm$0.04] \\
\hline
\end{tabular}
\end{table*}

\subsection{The central star}
The central star's position and magnitudes, as measured from the IPHAS
data, are $RA(2000)$= 01$^h$ 25$^m$ 07.93$^s$; $Dec(2000$)= +63$^o$
56$\arcmin$ 52.7$\arcsec$; $r'$=18.12$\pm$0.02, and $i'$=17.75$\pm$0.02,
respectively.
The $r'$ magnitude is likely to include nebular emission (c.f.
Fig.~\ref{iphas}). The object is an infrared source (2MASS: \cite{cutri})
with $J$=16.0$\pm$0.1; $H$=15.4$\pm$0.1 and $K_s$=14.45$\pm$
0.09. Assuming it suffers the same extinction as the nebular gas
(\chb\ =2.0, implying $A_V$= 4.1 mag) the dereddened colours
$(J-H)_0$=0.17 and $(H-K_s)_0$=0.72 locate the object among the bulk
of Galactic PNe nuclei and separate it convincingly from main-sequence stars,
Mira variables and from the loci of thermal dust emitters
(\cite{larios}; see also \cite{patricia}). The object is also located in the
PNe region of the near-infrared diagnostic diagram by \cite{schmeja}.
According to Whitelock (1985), the $(J-H)_0$ vs. $(H-K_s)_0$ colours
are due to a combination of gas
and dust thermal emission, line emission (e.g. HeI at 1.08 $\mu$m and H$_2$ at
2.12 $\mu$m) and photospheric emission from the central star. This makes it
very difficult to derive physical information from the location of a
particular object, unless one has detailed knowledge of its radio,
infrared and optical properties.  The optical spectrum of the star has
been extracted from the WHT-ISIS data after subtracting the nearby
nebular emission from two adjacent zones of 1.4$\times$0.9
arcsec${^2}$ located symmetrically at 2.6$\arcsec$ from the star, so
that nebular lines are optimally supressed.  The resulting stellar
continuum is very red and almost featureless.

\begin{figure}
\includegraphics[angle=-90,width=8.8cm]{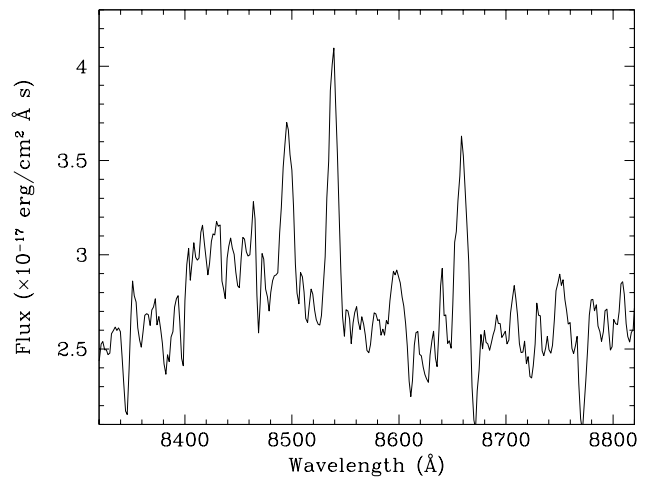}
\caption{CaII 8498, 8542 and 8662 {\AA}  emission lines at the central star.
The nebular spectrum has been subtracted. The Pa14 and Pa12 HI lines
at 8598 {\AA} and 8750 {\AA}, respectively, are barely detected.}
\label{CaII}
\end{figure}

The central star shows a very strong \ha\ emission ($I_{H\alpha}$= 2.5
10$^{-15}$ erg cm$^{-2}$ s$^{-1}$, integrated flux over an area of
1.4$\times$0.9 arcsec${^2}$) with a deconvolved gaussian $FWHM$ of 160
km~s$^{-1}$. The much fainter CaII lines at 8498, 8542 and
8662 {\AA}  (Fig.~\ref{CaII}) show a $FWHM$ velocity of
$\sim$200 km~s$^{-1}$.  Their dereddened
intensities are 7.1, 9.8 and 8.0 10$^{-18}$ erg cm$^{-2}$ s$^{-1}$,
respectively (we have neglected the contribution of the three Paschen
lines, Pa13, Pa15 and Pa16, blended with the CaII lines at our
spectral resolution, given the faintness of the nearby Pa12, Pa14 and
Pa17 lines; see Fig.~\ref{CaII}). The observed width of the CaII
lines is therefore similar to that of \ha , suggesting that both
originate in the same region around the star. The intensity ratio for
the CaII lines, 1.0:1.4:1.1, is very far from the expected ratio for
optically thin emission, 1:9:5. This is common for objects
where the triplet is in emission, and it indicates a high optical depth
(\cite{monica}). As in the case of the core of the PN He 2-428
discussed by the latter authors, the forbidden [CaII] lines at 7291
and 7324 {\AA} are undetected at the core of IPHAS PN-1,
indicating line quenching at very high densities ($N_e>$10$^{10}$
cm$^{-3}$) as would occur in either a circumnuclear disc or a very
dense circumstellar nebula.

\subsection{Spatio-kinematical modelling}
The [NII] $\lambda$ 6583 {\AA} data for $P.A.$= 0$^{\circ}$ described in
Section 2.3.1 (Fig.~\ref{spe}) show two maxima separated by 4.5\arcsec
and 4.6 km~s$^{-1}$. They correspond to the N and S sections of the inner
ring, whose projected expansion velocity is therefore 2.3 km~s$^{-1}$.
If it is assumed this ring has a circular shape, its observed dimensions
(Sect. 2.1) would imply an inclination (the angle between the symmetry axis
of the ring and the line of sight) of $\sim$68$^{\circ}$.

\begin{figure}
    \centerline{
   \epsfysize=5.0cm
    \epsffile{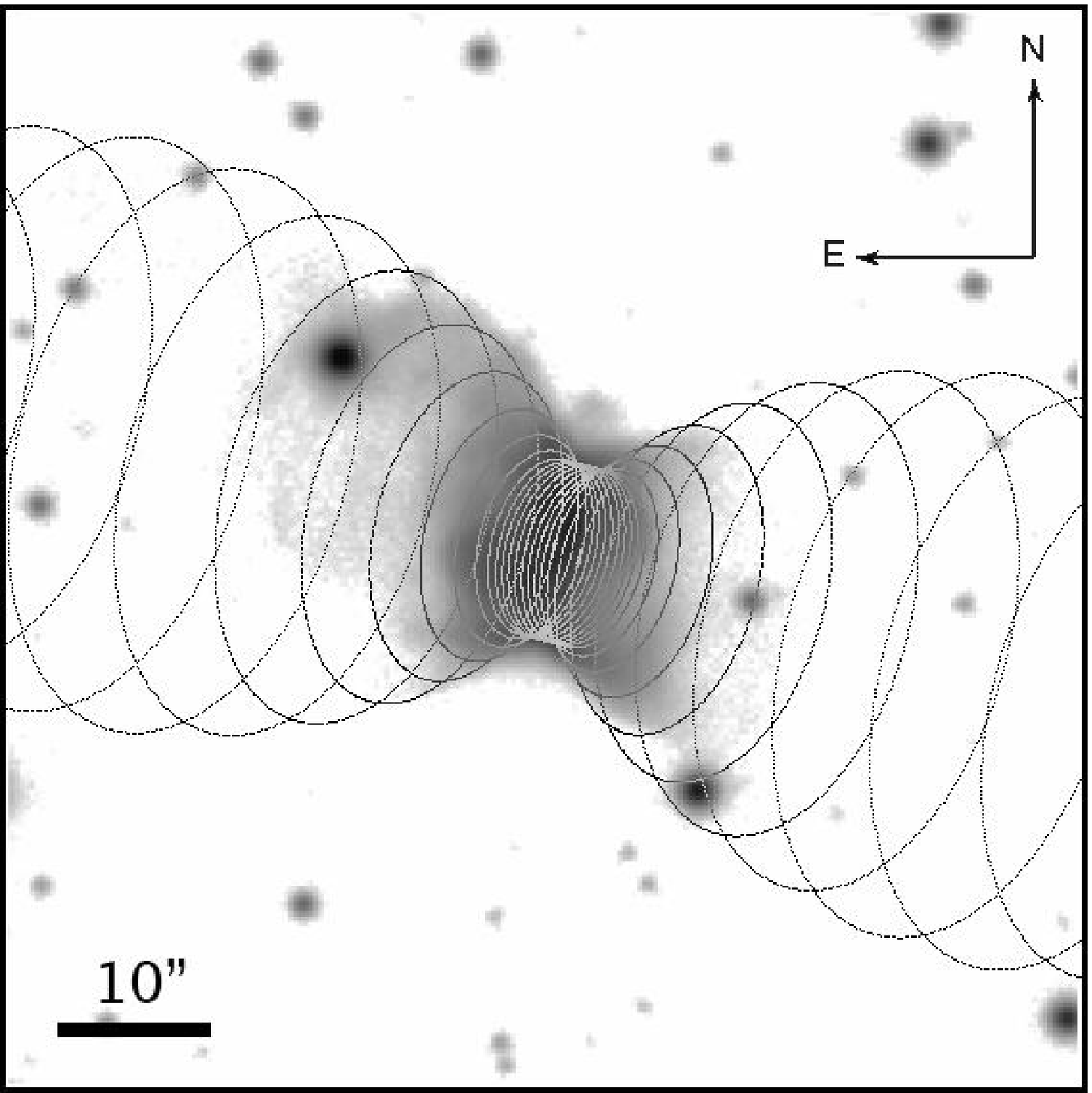}
    \epsfysize=5.0cm
    \epsffile{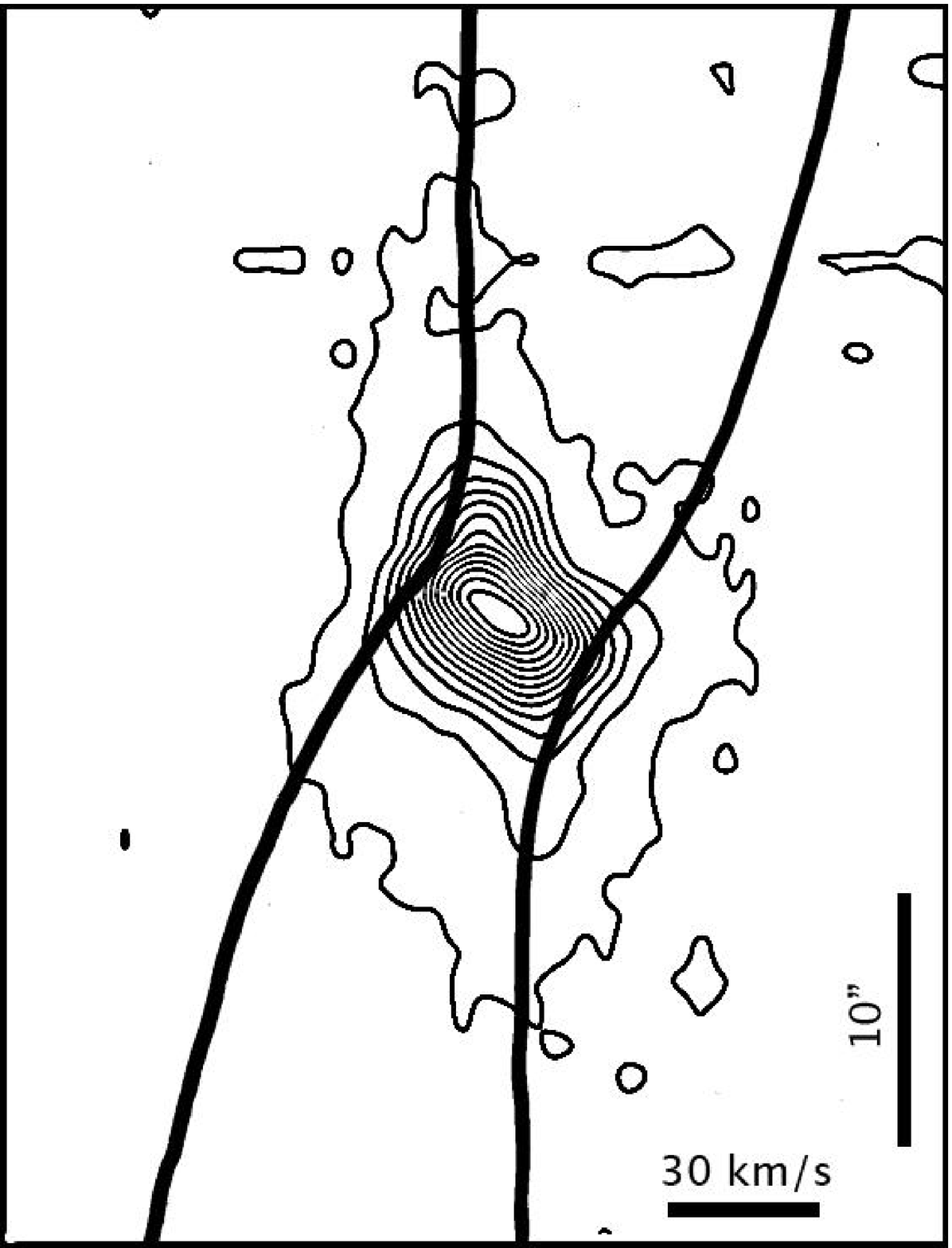}
    }
  \caption{({\it Left}) \ha +\nii\ image with superimposed
ellipses representing circles inscribed on the walls of the geometrical shape
used in the modelling. ({\it Right}) Observed [NII] position-velocity data for
$P.A.$= 57$^{\circ}$ (isocontours) together with corresponding modelling results
(thick lines).}
 \label{model}
\end{figure}

A simple spatio-kinematical model was developed to fit the velocity data at
$P.A.$= 57$^{\circ}$ (i.e. close to the symmetry axis of the inner nebula) and
the shape of the central waist and bright lobes (Fig.~\ref{sketch}). A
geometrical description following \cite{solf} was used, where the space
velocity of each gas particle is proportional to its distance from the central
star; this produces a self-similar expansion in a so-called ``Hubble-like''
flow (\cite{corradi2}). The model also assumes axial symmetry and radial
streamlines of gas. The resulting two-dimensional model is scaled to fit the
size of the object, rotated into three dimensions about its symmetry axis and
inclined to the plane of the sky to allow direct comparison with the image and
spectrum. Synthetic geometrical shapes and velocity-position plots are
generated and compared to the images and echelle spectrum. The
model fit to the data is carried out visually after allowing the kinematical
and geometrical parameters to vary over a large range of values; details of a
similar modelling applied to the planetary nebula Mz 3 can be found in
\cite{miguel}. Results for the best fit model for IPHAS PN-1 are presented in
Fig.~\ref{model}.

An inclination angle of 55$\pm$7$^{\circ}$ with respect to the line of
sight was found for the axis of symmetry of the lobes, i.e. somewhat  smaller
than the inclination measured for the ring (68$^{\circ}$). The waist expands
at a velocity of 11$\pm$3 km~s$^{-1}$.  The kinematic age of the nebula --
less constrained by the modelling -- is $\sim$2900 yr~kpc$^{-1}$, with a large
uncertainty spanning from 2000 to 4500 yr~kpc$^{-1}$. Assuming the above
mentioned Hubble-like expansion pattern, the lobe velocity at the poles would
be as high as ~220 km~s$^{-1}$, a figure that can be tested by deeper
observations able to reach the faint extremes of the lobes.

\subsection{Distance}
The object is located close to the Galactic plane at a favourable
longitude ($l$=126\gr.6) for obtaining an estimate of its distance via its
systemic radial velocity, by assuming that it participates in the
general circular rotation around the Galactic centre.  The
observed velocity of the nebula was measured from the higher-resolution
echelle observations of the
\ha\  and \nii\  lines in its central regions.  This was found to be
$-$83.5$\pm$1.5~\kms, which translates to $-$71.1$\pm$1.5~\kms\ when
corrected to the Local Standard of Rest. On adopting a standard Galactic
rotation curve (with a solar galactocentric distance of 8~kpc and a
circular velocity of 220~\kms), the kinematical distance of the nebula
is computed to be 7.0$^{+4.5}_{-3.0}$~kpc. The relatively large error
that we quote includes the error in the systemic velocity and, more
important, an estimate of the velocity dispersion along the line of
sight of all relatively young stars (age $\le$3~Gyr, consistent with
the hypothesis of a massive PN progenitor, see Sect. 4) located at a
given galactocentric distance, assuming the same values as for the
velocity dispersion ellipsoid in the solar neighbourhood
(\cite{nordstrom}).  This distance is only an estimate, as it comes
from a simplified treatment of the dynamics of our Galaxy, but clearly
points to a very large galactocentric distance of the nebula, namely
$D_{GC}$=13.4$^{+4.1}_{-2.5}$~kpc.

\begin{figure}
\includegraphics[angle=-90,width=8.8cm]{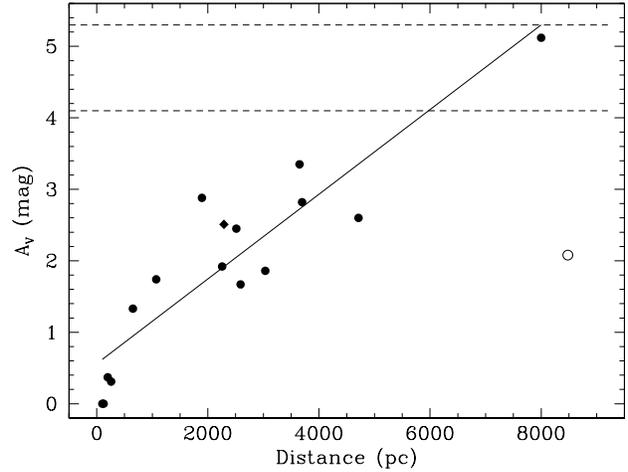}
\caption{Extinction-distance diagram for the field around IPHAS PN-1. The trend
line is based on a least squares fit to the field stars with best quality
data.   The upper dashed line shows the asymptotic reddening in this direction
and the lower dashed line is the reddening determined from the Balmer
decrement for the PN (see text).}
\label{EDR}
\end{figure}

The distance can also be obtained from an extinction-distance relationship
(e.g. \cite{kl}) for the field.  The SIMBAD astronomical database was used to
extract all available photometric and spectroscopic information for stars
within 60\arcmin\  of the PN.  Spectroscopic parallaxes were determined after
adopting the intrinsic colours and absolute magnitudes for each spectral class
from \cite{sk}.  Fig.~\ref{EDR} shows the trend, based on a linear
least-squares fit to the early-type (OBA) stars in the field with best-quality
data.  The upper dashed line shows the asymptotic reddening in this direction
(\cite{schlegel}) and the lower line is the reddening determined from the
Balmer decrement for the PN.

A monotonic relation of reddening versus distance is seen for this field,
although with considerable scatter.  Using the observed reddening to the PN
($A_{V}$=4.1$\pm$0.10), we obtain a distance of $D$=5.9$\pm 2.2$~kpc.  The
error on the distance is estimated following the approach of Kaler \& Lutz
(1985), and is probably optimistic as it does not account for any internal
reddening in the nebula (if present, $D$ is overestimated).  The diamond in
Fig.~\ref{EDR} is the intermediate-age open cluster NGC~559 (\cite{ann}),
while the point at $D$=8~kpc is the high-mass X-ray binary V635~Cas
(\cite{negueruela}). The point at lower right (open circle) is the B8 Ib star
NGC 559\#14 (\cite{lindoff}), that is not actually a member star of NGC 559.
The adopted
spectral type is from \cite{sowell} which places it far off the observed
trend, and so it is excluded from the fit.  However if the star is a bright
giant (luminosity class II) rather than a Ib supergiant, the star would fall
on the trend for the field.

Another distance estimation uses the empirical H$\alpha$ surface
brightness -- radius relation which has been recently calibrated with PNe
having accurate distances determined using a primary method (\cite{frew}; an
early version of this relation is presented by \cite{pierce}). The
technique requires an accurate
determination of the dimensions of the main body of a PN, excluding any outer
halos. For round and elliptical PNe, the adopted dimensions (used to
calculate the mean surface brightness) are easily estimated.  However, for
bipolar nebulae, estimating the size is more problematic, so the surface
brightnesses have a greater than average error.  The main body of IPHAS PN-1
was roughly fitted by an ellipse to get dimensions of
21$\arcsec\times$12\arcsec, and, using the total integrated \ha\  flux from
Sect. 3, and the relation of Frew et al. (2006) for a subset of well-studied
bipolar PNe, a distance, $D$=5.8$\pm$1.7~kpc is estimated.  This is in
excellent agreement with both the extinction distance and the kinematic
determination. In the following we will adopt the kinematical values, namely
$D$=7.0~kpc and $D_{GC}$=13.4~kpc.

The object is therefore located just outside the most
distant spiral arm in that direction, the Perseus+1 arm (\cite{vallee}) where
current star formation activity is evidenced by the presence of young
open clusters, OB associations, and HII regions (\cite{kw}).
Assuming for the whole nebula the inclination of 55$^{\circ}$ measured for the
inner regions (Sect. 3.3), a size of 4.3 pc is derived for the Eastern lobe,
making IPHAS PN-1 one of the largest PNe known. Its age is also very large:
using the expansion velocity determined in Sect. 3.3, we obtain $\sim$20000
yr for the nebula. The mass of the nebula can be estimated using the total
\ha\ flux calculated above, and results in a nebular mass of $\sim$0.05 \sm ,
i.e. towards the lower end of typical PNe masses (c.f. \cite{pottasch})
and similar to He 2-104 and other bipolar nebulae believed to be produced
by binary (or symbiotic) stars (\cite{cys}).

\subsection{Binarity}

It appears that the exciting star of IPHAS PN-1 belongs
to a binary system: at the assumed distance and reddening, its observed $i'$,
$J$ and $H$ magnitudes (where nebular contamination is expected to be minor;
Sect. 3.2) imply absolute magnitudes $M_{i^{\prime}}$= +0.9, $M_J$= +0.6,
and $M_H$=+0.4. The absolute $V$ magnitudes for the hot
central stars of evolved bipolar PNe (of which this is an example) range from
$M_{V}$ = +5.0 to +7.5 (\cite {peter}) equivalent to $M_{i^{\prime}} \approx$
+5.3 to +7.8, i.e. more than 4 mag fainter than observed. The
discrepancy remains even if the distance is reduced to our lower limit of
4~kpc. Since the star appears almost perfectly at the centroid of the
nebula, it is very unlikely that there is a line-of-sight projection.

We can use the implied $(J-H)_0$ and $(i'-J)_0$ colours of the nucleus
to take this discussion further.  We have derived $(J-H)_0 \simeq 0.2$,
$(i'-J)_0 \simeq 0.3$: these roughly match expectation for a mid-F star,
irrespective of luminosity class (cf. the synthetic photometry tabulated by
Pickles 1998, giving $J-H = 0.2-0.3$ and  $I_C-J \simeq 0.3$ at F5, for
luminosity classes V-III).   But it is clear that at a distance of 7~kpc, the
mid-F companion would have to be of luminosity class III ($M_V = 1.6$ is given
by Schmidt-Kaler 1982 for F5 III, c.f. $\sim 1.5$ implied for the central
star in this case).  At the lower limiting distance of 4~kpc the luminosity
falls to a value not so far above that of a dwarf star ($M_V = 3.5$ from
Schmidt-Kaler 1982, versus $\sim 2.7$ here, at the nearer distance).  A mid-F
companion is thus the simplest option, and would not require much of a
contribution to the red/NIR flux from accretion.  In this circumstance, the
source of the observed CaII IR triplet emission can be either irradiation of
the companion, or a circumnuclear disc in which accretion at rates typical of
high-state cataclysmic variables ($M_V \sim 4$, and neutral colour,
Warner 1995) would not be detected.  On the other hand, if e.g. NIR
spectroscopy were to fail to confirm an F-star photosphere, then the more
exotic scenario of light from both an evolved
later-type companion and a relatively luminous accretion component would have
to be considered (for all distances in excess of the 4~kpc minimum): such a
system could resemble the old nova, and extreme dwarf nova, GK Per ($M_V \sim
1.6$ in outburst).  In either case, there is no escaping the need to model
this PN nucleus as a binary, and there is a significant likelihood that the
binary is a relatively close one, with a period on the order of a day, so as
to permit either significant irradation of the companion or a semi-detached
configuration leading to mass transfer and accretion.

If the nuclear binary period is not short, dense circumstellar gas within a
passive disk or similar remains a necessity as discussed in Section 3.2.
High resolution imaging (e.g with $HST$) might reveal this.

\section{Discussion}
The group of quadrupolar PNe is composed of only eight objects: five
nebulae originally identified by Manchado et al. (1996; K 3-24, M
1-75, M 2-46, M 3-28, and M 4-14), one (NGC 6881) by Guerrero \&
Manchado (1998), a further one (NGC 4361) by \cite{muthu}, plus NGC
2440 (\cite{luis}). All of them show an enhanced waist and two pairs
of differently-aligned bipolar lobes, except in
NGC 4361 whose nature as a quadrupolar PN is doubtful. IPHAS
PN-1 is morphologically very similar to M 2-46 and M 1-75
but also to the bipolar PNe A~79 and He 2-248 -- the latter are rather
evolved nebulae, excited by binary stars (\cite{monica}).

The origin of quadrupolar nebulae is important for the theories of PNe
formation and evolution because they pose a strong challenge to the
paradigm of the Generalized Interacting Stellar Winds
(GISW; \cite{balick}). GISW is not able to account for the formation of
quadrupolar (nor multipolar and point-symmetric) nebulae and further
physical processes such as external torques of a close binary companion or
strong magnetic fields in the wind ejecta are required.
Furthermore, there is an unsolved evolutionary problem in
that bipolar, quadrupolar and multipolar morphologies are in the majority among
proto-PNe and young PNe (\cite{sahai98}), whereas round and elliptical
geometries are, on the contrary, much more frequent both among PNe in general
(\cite{manchado04}) and among their precursor AGB shells
(c.f. \cite{sahai04}). The process that transforms a spherical AGB envelope
into an aspherical (but axisymmetric) object, like a quadrupolar PN,
represents a major challenge for the theories of
post-main-sequence evolution (\cite{sahai98}).

Models proposed for quadrupolar PNe
are varied: a precessing binary system ejecting two bipolar shells
(\cite{manchado}); an helicoidal precessing jet (Guerrero \& Manchado,
1998); magnetized, misaligned winds from a star and disc system
(\cite{blackman}); and the precession of warped discs (\cite{livio};
\cite{icke}). The one item in common with all models is the
required presence of a binary system, although no evidence exists for
that in the known sample of quadrupolars.  Here, the evidence for
binarity is compelling.  Whether the binary is actually experiencing mass
transfer is not so clear at this point. Nevertheless, as discussed in Sects.
3.2 and 3.5 above, there remains the possibility also of a very dense
circumnuclear shell or disc in IPHAS PN-1, in spite of the fact that the
nebula is rather old:  it is hard, otherwise, to understand the widths
and relative strengths of the CaII IR triplet components.

The chemistry of IPHAS PN-1, presented in Sect. 3.1,
shows that it is a Type I PN, with $He/H$=0.13$\pm$0.01 and
$N/O$=1.8$\pm$0.6. The O abundance is remarkably low (Table~2, but
notice the large associated error caused by the uncertain $T_e$[O{\sc
iii}]) and similar to the PNe with high-mass progenitors discussed by
\cite{marigo}).
However, we note that the Kingsburgh \& Barlow (1994; KB94 in the following)
{\it icf} scheme may break down for extreme
Type~I nebulae such as IPHAS PN-1 and NGC~6302. Heavy
elements in the latter nebula are spread across a much wider range of ion
stages than for `normal' PNe. Using just the O$^{+}$ and O$^{2+}$
abundances for NGC~6302 from Table~9 of \cite{tsamis}, equation A7
of KB94 predicts an {\it icf}(O) of 1.62. However, the O~{\sc iv}]
$\lambda$1401-based abundance of O$^{3+}$ was alone found by Tsamis et al.
to be 82\% of the O$^{+}$ + O$^{2+}$ abundance, with the {\it icf}(O) for
stages higher than O$^{3+}$ estimated by them to be 1.37, using equation A9 of
KB94. As a result, Tsamis et al. obtained an overall oxygen abundance for
NGC~6302 that was a factor of 1.6 larger than would have been estimated
based just on the O$^{+}$ and O$^{2+}$ abundances. Even this may be an
underestimate, since NGC~6302 exhibits strong near-IR coronal line
emission from highly ionized species (e.g. [Si~{\sc vii}] 2.48~$\mu$m and
[Si~{\sc ix}] 3.93~$\mu$m; see \cite{casassus}), suggesting that
highly ionized stages of oxygen may also be present. Since
IPHAS PN-1 has a similar morphology and He~{\sc ii}/H$\beta$
ratio to NGC~6302 (\cite{matsuura}), the $O/H$ abundance ratio listed in
Table~2 could be a
lower limit. Infrared spectroscopy of the [O~{\sc iv}] 25.89-$\mu$m line,
as well as spectra covering the near-IR coronal lines, could be useful in
helping to constrain the oxygen abundance.

From Figs. 11 and 12 of \cite{marigo}; Ma03 in the following), a progenitor
mass of 2.5$\to$3 \sm\ would be estimated for IPHAS PN-1. But it is
important to note that the object
occupies an anomalous zone in the $He/H$ vs. $N/O$ diagram, where models
with initial solar metallicity and masses up to 5 \sm\ fall short by a
factor $>$3 in reproducing the observed $N/O$, while the models with
initial LMC metallicity overestimate $He/H$ by $>$25{\%}, a figure well
outside the $He/H$ error bar. Interestingly, IPHAS PN-1 is
not alone in this ``forbidden" zone of the $He/H$ vs. $N/O$ diagram: two
other quadrupolar PNe are there, NGC 2440 (Ma03) and M1-75
(\cite{perinotto04}). In addition, two Type I PNe from the Ma03
article, NGC 5315 and NGC 6537 also lie in the same area of this
diagram. In fact, the group of Type I PNe studied by KB94 has
average values of $He/H$=0.13 and $N/O$=1.2, i.e. they are also located in
roughly the same area of the Ma03 diagram, pointing to a lack of
plausible models for this kind of object (notice, however, that the
Ma03 models do reproduce the abundances of the most extreme Type I bipolars
in their sample, i.e. those with $He/H\geq$0.15). With these considerations,
we estimate that the progenitor of IPHAS PN-1 was
an intermediate mass star (2.5$\to$3 {\sm}) with solar
metallicity or lower.

Let us compare now IPHAS PN-1 with the other known
quadrupolar PNe. The first thing to note is that the quadrupolars are
not a chemically homogeneous group: for the six PNe where adequate
data exist, two are of Type I (M 1-75 and NGC 2440), two of Type II (M
4-14 and NGC 6881), one doubtful object (M 2-46) and one Type IV
(Halo) PN, NGC 4361 (but see above about its unclear
classification). (References for the chemical abundances for these
objects can be found in \cite{tpp};
\cite{perinotto91}; \cite{koppen}; Ma03; and
\cite{perinotto04}). IPHAS PN-1 is the most extreme Type I
nebula (in He abundance and $N/O$ ratio) of the known quadrupolars.

Table~2 compares IPHAS PN-1 with two Type I samples: a) the group of 11
objects measured by KB94, all but one located at $D_{GC}{\leq}$ 11~kpc, and b)
three PNe from \cite{costa} having the largest $D_{GC}$ in their sample: M
1-18, located at 10.9~kpc, M 3-3, at 12.4~kpc, and M 3-2, at 14.1~kpc (note
that these authors adopt a solar galactocentric distance of 7.5~kpc).  Subject
to the caveat that unaccounted-for high ion stages could be present (see
above), the abundances for IPHAS PN-1 in Table~2 are generally lower than for
PNe in both the a) and b) samples, indicating a lower metallicity progenitor,
and therefore in qualitative agreement with the large $D_{GC}$ of 13.4~kpc
estimated above. \cite{costa} present a detailed study of chemical abundances
for PNe towards the galactic anticentre. Their main conclusion, a flattening
of the $O/H$ gradient at large ($\ga $11~kpc) galactocentric distances, is
however hampered by the large observational dispersion (0.3 dex) for those
distant objects. Clearly, more objects and better determined abundances and
distances are needed to confirm this flattening. IPHAS PN-1 is an important
PN in this respect: its $O/H$ is the second lowest (after K 3-68; c.f. Fig. 4
in Costa et al. 2004) and its $D_{GC}$ one of the largest for which reliable
data exist\footnote {An additional very distant PN has been recently found
from the IPHAS survey: it is a Type II object with $O/H$=8.5 and it is very
likely located beyond 14 kpc from the Galactic centre (\cite{mampaso}).}.  In
fact, the low $O/H$ measured for IPHAS PN-1, even allowing for an {\it icf}(O)
enhanced from 1.5 (Table~2) to 2.4 (were it identical to NGC 6302; see above),
is consistent with the galactic gradients found from PNe (--0.05
dex~kpc$^{-1}$; Costa et al. 2004) and early B stars (--0.07 dex~kpc$^{-1}$;
\cite{rolleston}) at the galactocentric distance determined for the object.
This lends support to models
where the oxygen abundance gradient has a constant slope (\cite{henry}).

\section{Conclusions}
The INT Photometric H$\alpha$ Survey (IPHAS) is currently mapping the Northern
Galactic Plane at unprecedented depth and spatial resolution.  Hundreds of new
planetary nebulae are awaiting discovery in the IPHAS photometric catalogue
and its combined mosaic images. Here we have presented a morphological and
physico-chemical study of the first PN discovered, IPHASX
J012507.9+635652. This is an unusual nebula composed of a compact elliptical
ring, inner lobes and waist, and faint outer bipolar lobes extending up to
more than 100$''$ from the central star.  The source likely belongs to the
group of quadrupolar nebulae, with only eight other members known so far, and
further high spatial-resolution observations, and high-resolution spectroscopy
of the nebula are desirable in order to unambiguously determine its
quadrupolar nature.  Broad H$\alpha$ and CaII emission lines are detected from
the central star, indicating the presence of very high densities
($N_e>$10$^{10}$ cm$^{-3}$) from a thick circumstellar shell or disc. This,
together with the photometric properties of the central star, make unavoidable
the presence of a binary nucleus, the first evidence for this among the
quadrupolars. Whether there is an active or a passive disk (or some other
geometric structure) cannot be decided with the data at hand.  The inner
nebula is heavily reddened ($A_V$= 4.1 mag), with low densities ($N_e$[S{\sc
ii}] = 390$\pm$40 cm$^{-3}$) and rather high temperatures ($T_e$[O{\sc iii}] =
14100$\pm$2800 K, $T_e$[N{\sc ii}] = 12800$\pm$1000 K). The chemistry is
typical of extreme Type I PNe ($He/H$ and $N/O$ attaining very large values),
but it shows very low Oxygen, Neon, Sulphur, and Argon abundances, all being
consistent with an intermediate-mass progenitor formed in a low metallicity
environment.  The distance to IPHASX J012507.9+635652 is probably very large:
a kinematic distance of 7.0$^{+4.5}_{-3.0}$~kpc is derived from the \ha\ and
\nii\ radial velocities, yielding a huge galactocentric distance of
13.4$^{+4.1}_{-2.5}$~kpc. This makes IPHAS J012507.9+635652 a rare and most
valuable probe for chemical studies in the outer Galaxy.

\begin{acknowledgements}
This work is based on observations made with the INT and WHT telescopes
operated on the island of La Palma
by the Isaac Newton Group (ING) in the Spanish Observatorio del Roque de los
Muchachos.
INT and WHT observations were made during service time, and the excellent
support from the ING staff is
sincerely ackowledged.  We thank J.A. L\'opez, M. Richer, and  H. Riesgo for
kindly acquiring the MESCAL spectra for us, and
M. Santander Garc{\'{\i}}a for the use of his spatio-kinematical programmes.
A.M., R.L.M.C., K.V., E.R.R.F, and P.L. thank funding
from the Spanish AYA2002-0883 grant. Finally, A.M. acknowledges the hospitality
of the Instituto Nacional de
Astrof{\'{\i}}sica, \'Optica y Electr\'onica (INAOE, Puebla, M\'exico) and the
Spanish M.E.C (grant PR 2004-0598) during his
sabbatical leave in M\'exico.

\end{acknowledgements}

\end{document}